\documentclass[journal=jpclcd,manuscript=article,layout=preprint]{achemso}

\usepackage{amsfonts}
\usepackage{bm}
\usepackage{setspace}
\usepackage{graphicx}
\usepackage{color}
\usepackage{verbatim}
\usepackage{float}
\usepackage{array}
\usepackage{booktabs}
\usepackage{amsmath}
\usepackage{amssymb}
\usepackage{graphicx}
\usepackage[version=3]{mhchem}
\usepackage[T1]{fontenc}
\usepackage{subfigure}
\usepackage{subfig}
\usepackage{xr}
\usepackage{threeparttablex}
\usepackage{xcolor}
\usepackage{enumitem}
\usepackage{soul}
\usepackage{relsize}
\usepackage{listings} 
\usepackage[T1]{fontenc}
\usepackage{widetable}
\usepackage{multirow}
\usepackage{longtable}
\usepackage{threeparttable}
\usepackage{rotating}
\usepackage{booktabs}
\usepackage{pdflscape}
\usepackage{diagbox}
\usepackage{textcomp}

\definecolor{dkgreen}{rgb}{0,0.6,0}
\definecolor{gray}{rgb}{0.5,0.5,0.5}
\definecolor{mauve}{rgb}{0.58,0,0.82}
\newcommand{\dd}{\mathop{}\!\text{d}}

\newcommand{\lozero}[0]{\texttt{no localization}}
\newcommand{\loone}[0]{\texttt{localization1}}
\newcommand{\lotwo}[0]{\texttt{localization2}}
\newcommand{\kappazero}[0]{\texttt{curvature0}}
\newcommand{\kappaone}[0]{\texttt{curvature1}}
\newcommand{\kappatwo}[0]{\texttt{curvature2}}
\newcommand{\libsc}[0]{{\texttt{LibSC}}}
\newcommand{\Rplus}{\protect\hspace{-.1em}\protect\raisebox{.35ex}{\smaller{\smaller\textbf{+}}}}
\newcommand{\Cpp}[0]{\mbox{C\Rplus\Rplus}}
\newcommand{\Cpponeone}[0]{\mbox{C\Rplus\Rplus11}}
\newcommand{\psilosc}[0]{\texttt{psi4\_losc}}
\newcommand{\pyscflosc}[0]{\texttt{pyscf\_losc}}
\newcommand{\pyscf}[0]{\texttt{PySCF}}
\newcommand{\psipkg}[0]{\texttt{Psi4}}
\newcommand{\qmfd}[0]{\texttt{QM4D}}

\title{
\libsc{}: Library for Scaling Correction Methods
in Density Functional Theory
}

\author{Yuncai Mei}
\author{Jincheng Yu}
\author{Zehua Chen}
\author{Neil Qiang Su}
\affiliation{Department of Chemistry, Duke University, Durham,
    North Carolina 27708, USA}
\footnotetext[1]{
Y.M. and J.Y. contributed equally to this work and share the
first authorship.}

\author{Weitao Yang}
\email{weitao.yang@duke.edu}
\affiliation{Department of Chemistry, Duke University, Durham,
    North Carolina 27708, USA}
\alsoaffiliation{Department of Physics, Duke University, Durham,
    North Carolina 27708, USA}
\date{\today}

\makeatother

\begin{document}
\lstdefinelanguage{inp}
{morekeywords={import},
sensitive=false,
morecomment=[l]{\#}
}
\lstdefinestyle{numbers}
{numbers=left, stepnumber=1, numberstyle=\scriptsize, numbersep=10pt}


\begin{abstract}
In recent years, a series of scaling correction (SC) methods have
been developed in the Yang laboratory to reduce and eliminate the
delocalization error, which is an intrinsic and systematic error existing
in conventional density functional approximations (DFAs) within density
functional theory (DFT). Based on extensive numerical results, the
SC methods have been demonstrated to be capable of reducing the delocalization
error effectively and producing accurate descriptions for many critical
and challenging problems, including the fundamental gap, photoemission
spectroscopy, charge transfer excitations and polarizability. In the
development of SC methods, the SC methods were mainly implemented
in the \qmfd{} package that was developed in the Yang laboratory
for research development. 
The heavy dependency on the \qmfd{} package hinders
the SC methods from access by researchers for broad applications.
In this work, we developed a reliable and efficient implementation
, \libsc{} for the global scaling correction (GSC) method and the
localized orbital scaling correction (LOSC) method. \libsc{} will
serve as a light-weight and open-source library that can be easily
accessed by the quantum chemistry community. The implementation of
\libsc{} is carefully modularized to provide the essential functionalities
for conducting calculations of the SC methods. In addition, \libsc{}
provides simple and consistent interfaces to support multiple popular
programing languages, including C, \Cpp{} and Python. In addition to the
development of the library, we also integrated \libsc{} with two
popular and open-source quantum chemistry packages, the \psipkg{}
package and the \pyscf{} package, which provides immediate access
for general users to perform calculations with SC methods.
\end{abstract}
\clearpage{}
\begin{tocentry}
\includegraphics{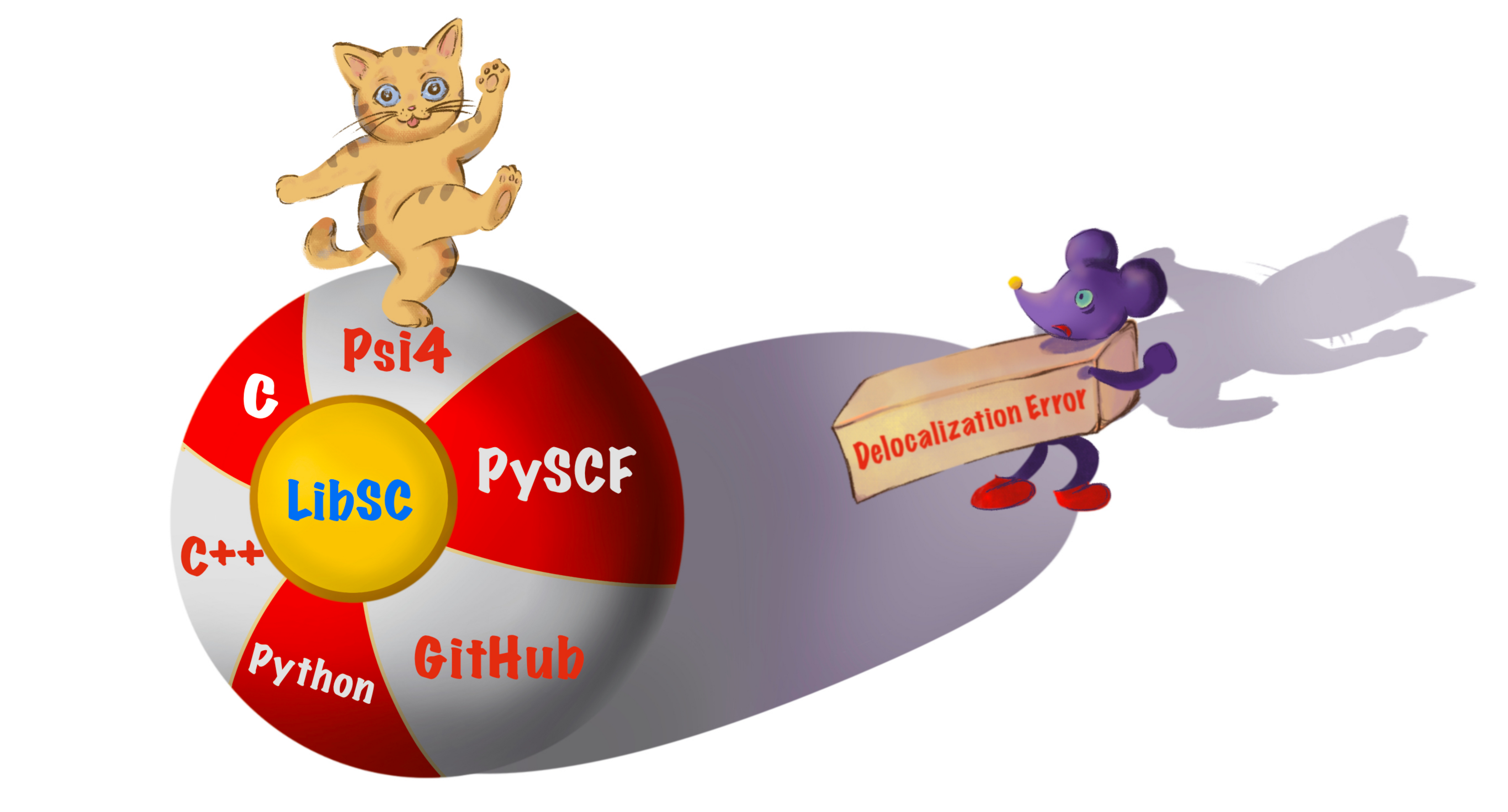}
\end{tocentry}

\section{Introduction}

Density functional theory (DFT) \cite{1964hohenbergB871,1965kohnA1138,1989parr}
has been widely used nowadays to describe the electron structure of
matters in chemistry, physics and materials science. In the pursuit
of accurate predictions from theoretical simulations based on DFT,
developing accurate density functional approximations (DFAs) within
DFT has become an active research field in quantum chemistry and condensed
matter physics. During the last decades, conventional DFAs, including
the local density approximations (LDAs) \cite{1980vosko1211,1992perdew13249},
generalized gradient approximations (GGAs) \cite{1988becke3100,1988lee789,1996perdew3868}
and hybrid functionals \cite{becke1993new,1988lee789,1994stephens11627,1999adamo6170,1999ernzerhof5036},
have achieved great success. However, conventional DFAs involve the
intrinsic and systematic delocalization error \cite{2008mori-sanchez146401,2008cohen115123,2008cohen794,2012cohen320},
which is the underlying challenge for many critical applications \cite{1998zhang2608,2006dutoi233,2006mori-sanchez201102,2006ruzsinszky194112,2007vydrov154109,2008cohen794,2008mori-sanchez146401,2012cohen320,2012zheng214106}.

To reduce the systematic delocalization error, many approaches have
been developed, including range-separated functionals \cite{1995savin332,1996savin357,2001iikura3544,2004yanai57,2006vydrov234109,2008chai6620, baer2010tuned},
self-interaction error corrected functionals\cite{1981perdew5079,2006mori-sanchez201102,2006mori-sanchez091102,2008perdew052513,2014schmidt14367,2014pederson121103,2016schmidt165120,2017yang052505},
Koopmans-compliant functionals, \cite{2014borghi075135,2019colonna1914}
and generalized transition state methods \cite{2005anisimov075125}
and related developments \cite{ma2016using}.

In addition to these aforementioned approaches, a series of scaling correction
(SC) methods \cite{2011zheng026403a,2015li053001a,2018li215,2020su1535,2020mei10277,2021mei10358},
including the global scaling correction (GSC) \cite{2011zheng026403a,2021mei10358},
the local scaling correction (LSC) \cite{2015li053001a} and the localized
orbital scaling correction (LOSC) \cite{2018li215,2020su1535,2020mei10277}
methods, have been developed in the Yang laboratory to tackle the
delocalization error in conventional DFAs. With extensive numerical
results \cite{2011zheng026403a,2015li053001a,2018li215,2018li215,2019mei673,2019mei144109,2019mei2545,2020yangc,2021mei054302},
these SC methods have been demonstrated to be capable of reducing
the delocalization error effectively and producing accurate descriptions
for many critical and challenging problems, including the fundamental
gap \cite{2011zheng026403a,2018li215,2019mei673}, photoemission spectroscopy
\cite{2018li215,2019mei673,2020yangc}, photoexcitation energies \cite{2019mei673,2019mei144109,2019mei2545}
and polarizability \cite{2015li053001a,2021mei054302}. Therefore,
a reliable and stable implementation for the SC methods can be very
beneficial and meaningful to the electronic structure theory community,
which helps to promote DFT with commonly used DFAs and with the SC methods 
for broader applications.

However, along the development of SC methods, the implementation was
mainly developed in the \qmfd{} package \cite{qm4d}, which is an
in-house quantum chemistry package in the Yang laboratory for research
development. 
%
In this work, we developed a reliable and stable implementation for
the GSC and LOSC methods in \libsc{}, which will serve as a light-weight
and open-source library to provide the essential functionalities for
conducting calculations of the SC methods. With the simple and consistent
interface to support multiple popular programing languages, including
C, \Cpp{} and Python, we aim to provide future \libsc{} users
great flexibility to implement the SC methods in different quantum
chemistry packages of their choices in a short development cycle.
In addition to the development of the library, we also integrated \libsc{}
with two popular and open-source quantum chemistry packages, the \psipkg{}
package \cite{2020smith184108} and the \pyscf{} package \cite{2018sune1340},
which provides immediate access for researchers to perform calculations
of SC methods easily. We will describe the philosophy and methodology
of the design for \libsc{} and show its applications.

\section{Theoretical Background}

We start with a brief review for the key concept of the delocalization
error \cite{2008mori-sanchez146401,2008cohen115123,2008cohen794,2012cohen320},
which is the central problem that the SC methods are designed to solve.
The delocalization error characterizes the incorrect behavior of conventional
DFAs compared to the exact functional in DFT, and it can be understood
from the perspective of systems with fractional number of electrons.
According to the Perdew-Parr-Levy-Balduz (PPLB) condition \cite{1982perdew1694,2000yang5175,2001zhang348},
the exact total energy $E(N)$, as a function of electron number,
should be piecewise linear between any two adjacent integer points.
Critically, the manifestation of the delocalization error has been
shown to be size-dependent \cite{2008mori-sanchez146401}. For small
systems, conventional DFAs usually well predict total energies for
integer systems. However, conventional DFAs severely underestimate
the total energies for fractional systems \cite{2008mori-sanchez146401,2008cohen115123}.
Such underestimation from conventional DFAs leads to a convex $E(N)$
curve, which is the manifestation of the delocalization error for
small systems. For large systems, the behavior is different. The deviation
of the $E(N)$ from the linearity condition decreases when the size
of the system becomes larger, and vanishes at the bulk limit \cite{2008mori-sanchez146401}.
However, the delocalization error manifests as the underestimation
for the total energies of integer systems with the addition or removal
of an electron, which produces the $E(N)$ curve with wrong slopes
at the bulk limit \cite{2008mori-sanchez146401}. The direct consequence
for the delocalization error is the large error in the prediction
of chemical potentials, which are first derivatives of the total energy
with respect to the electron number, $\frac{\partial E}{\partial N}\Big|_{\pm}$
from the two sides of the integer $N$. The chemical potentials 
$\frac{\partial E}{\partial N}\Big|_{-}$ and 
$\frac{\partial E}{\partial N}\Big|_{+}$ have been rigorously
proved to be the energy of the highest occupied molecular orbital
(HOMO) and the energy of the lowest unoccupied molecular orbital (LUMO)
respectively \cite{2008cohen115123} within the Kohn-Sham DFT,
in which the approximate exchange-correction energy is an explicit functional
of electron density, or the generalized Kohn-Sham DFT, in which the
approximate exchange-correction energy is an explicit functional of the first-order
density matrix.
Therefore, when the PPLB condition \cite{1982perdew1694}
is satisfied, HOMO and LUMO energies connect the first ionization
potential (IP) and the first electron affinity (EA) \cite{2008cohen115123},
which read 
\begin{align}
\epsilon_{{\rm {HOMO}}} & =\frac{\partial E}{\partial N}\Big|_{-}=E(N)-E(N-1)=-{\rm {IP},}\label{eq:homo}\\
\epsilon_{{\rm {LUMO}}} & =\frac{\partial E}{\partial N}\Big|_{+}=E(N+1)-E(N)=-{\rm {EA}.}\label{eq:lumo}
\end{align}
According to Eqs.~\ref{eq:homo} and \ref{eq:lumo}, the direct results
of the delocalization error are the underestimation of the first IP
from the HOMO energy and the overestimation of the first EA from the
LUMO energy, thus the drastic underestimation of the fundamental gap
from the HOMO-LUMO energy gap \cite{2008cohen115123}.

To reduce the delocalization error, the SC methods impose the PPLB
condition to associated DFAs either ``globally'' or ``locally'' to
construct total energy corrections and restore the linear behavior 
between integers. Specifically, the global scaling
correction (GSC) method \cite{2011zheng026403a} imposes the PPLB
condition globally through the canonical orbitals and their occupation
numbers. Within the GSC, the total energies of integer systems remain
the same as those from the parent DFA. The energy correction from the GSC
is constructed as the energy compensation to the corresponding linear
interpolation and it is effective for fractional systems only, which
is expressed as an addition to the total energy 
\begin{align}
\Delta_{{\rm {GSC}}}(N+n)=(1-n)E(N)+nE(N+1)-E(N+n).\label{eq:gsc_E}
\end{align}
Based on Eq.~\ref{eq:gsc_E}, the energy correction for GSC at the
second order is given as 
\begin{align}
\Delta_{{\rm {GSC}}}=\frac{1}{2}\sum_{p\sigma}\kappa_{p\sigma}(n_{p\sigma}-n_{p\sigma}^{2}),\label{eq:gsc_corr}
\end{align}
in which $n_{p\sigma}$ is the canonical orbital occupation number. Note that
the original consideration of HOMO and LUMO \cite{2011zheng026403a}
has been generalized to all the orbitals \cite{2020yangc,2021mei10358}.
The coefficient $\kappa_{p\sigma}$ in the original work of GSC \cite{2011zheng026403a}
is approximated explicitly as 
\begin{align}
\kappa_{p\sigma}=\int\frac{\rho_{p\sigma}(\mathbf{r})\rho_{p\sigma}(\mathbf{r'})}{|\mathbf{r}-\mathbf{r'}|}\dd\mathbf{r}\dd\mathbf{r'}-\frac{2\tau C_{\text{x}}}{3}\int\big[\rho_{p\sigma}(\mathbf{r})\big]^{\frac{4}{3}}\dd\mathbf{r},\label{eq:gsc_kappa}
\end{align}
in which the first term is the Coulomb interaction,
and the second term is attributed to the Slater exchange energy. The
parameters are chosen as $\tau=1$, $C_{{\rm {x}}}=\frac{3}{4}(\frac{6}{\pi})^{1/3}$,
and $\rho_{p\sigma}$ is the corresponding canonical orbital density
$|\psi_{p\sigma}|^{2}$. 
The exact coefficient $\kappa_{p\sigma}$ has been derived in a recent
development of GSC (GSC2) \cite{2021mei10358} and it is given as
the second order derivative of the total energy with respect to the
canonical orbital occupation number, $\kappa_{p\sigma}=\frac{\partial^{2}E}{\partial n_{p\sigma}^{2}}$,
which is a completely different and more sophisticated expression compared
to Eq.~\ref{eq:gsc_kappa}. 
Note that in the recent work from Xiao and coworkers \cite{2020yangc},
they also developed GSC to involve high order relaxation of orbitals
with respect to the canonical orbital occupation number and thus the energy
correction goes beyond the second order expression as shown in Eq.~\ref{eq:gsc_corr}.
The advantage of GSC2 is that it provides exact second order corrections
for any DFA \cite{2021mei10358}.

Note that the GSC preserves the total energies for integer systems
and only corrects fractional systems, meaning the contribution from
an orbital with an integer occupation ($n_{p\sigma}=0$ or $n_{p\sigma}=1$)
is zero. Therefore, the GSC is limited to reducing the delocalization
error for small and medium-sized systems. To treat the delocalization
error for large systems, where the violation of the PPLB condition is no
longer an issue and cannot be used as measure of the delocalization error
\cite{2008mori-sanchez146401},
the local scaling correction (LSC) \cite{2015li053001a}
method was developed later, which focuses on the local regions of
the molecular system and applies the corrections locally. Combining
the strategies used in GSC and LSC, the localized orbital scaling
correction (LOSC) \cite{2018li215,2020su1535,2020mei10277} was developed
to provide a general approach for the systematical elimination of
the delocalization error for both small and large systems. The key
idea in the LOSC is to change the canonical orbitals used in GSC to
the carefully designed localized orbitals, or ``orbitalets'', which
are one-electron orbitals that are localized both in space and in
energy and are unitary combinations of both the occupied and unoccupied 
orbitals. The localization used in the latest version of the LOSC (called
LOSC2) \cite{2020su1535} is defined as the minimization of the following
cost function 
\begin{equation}
    F^{\sigma}
    =(1-\gamma)\sum_{p}\Big(\langle\mathbf{r}^{2}\rangle_{p\sigma}-\langle\mathbf{r}\rangle_{p\sigma}^{2}\Big)+\gamma C\sum_{p}\Big(\langle H_{\sigma}^{2}\rangle_{p\sigma}-\langle H_{\sigma}\rangle_{p\sigma}^{2}\Big),,\label{eq:localization}
\end{equation}
where 
\begin{equation}
\langle X\rangle_{p\sigma}=\langle\phi_{p\sigma}|X|\phi_{p\sigma}\rangle,\quad X=\mathbf{r},\mathbf{r}^{2},H_{\sigma},H_{\sigma}^{2},
\end{equation}
\begin{equation}
\phi_{p\sigma}=\sum_{q}U_{pq}^{\sigma}\psi_{q\sigma},\label{eq:localization_end}
\end{equation}
$\sigma$ is the electron spin, $\gamma$ is set to be 0.707, $C$
is set to be 1000 a.u./\r{A}, $U_{pq}^{\sigma}$ is a unitary matrix, $\{\psi_{q\sigma}\}$
is the set of canonical orbitals, $\{\phi_{p\sigma}\}$ is the set
of orbitalets, $\mathbf{r}$ is the spacial operator, and $H_{\sigma}$
is the one-electron Hamiltonian of the parent DFA. 
Note that we use $\psi_{i\sigma}$ and $U_{pi}^{\sigma}$ to express 
the dependence on the orbitalets of the localization functional $F^{\sigma}$
in Eq.~\ref{eq:localization}.
According to Eqs.~\ref{eq:localization}-\ref{eq:localization_end},
the orbitalets can adaptively change between the canonical orbitals
and the localized orbitals, which are determined by the localization
for the system of interest at the given structure. According to Eq.~\ref{eq:localization_end},
one can apply an energy window to select COs to control the dimension
of the space for localization, thus improving the computational
efficiency of the localization. In practice, an energy window of {[}$-$30,
10{]} eV that covers most of the valence occupied COs and low-lying
virtual COs is applied as an approximate treatment to achieve more
efficient calculations.

With the application of orbitalets, the energy correction of the LOSC
is generalized from the GSC to be 
\begin{align}
\Delta_{{\rm {LOSC}}}=\frac{1}{2}\sum_{pq\sigma}\kappa_{pq\sigma}\lambda_{pq\sigma}(\delta_{pq}-\lambda_{pq\sigma}),\label{eq:losc_corr}
\end{align}
in which $\lambda_{pq\sigma}$ is the representation of density operator
under the orbitalets, namely $\lambda_{pq\sigma}=\langle\phi_{p\sigma}|\rho_{s}^{\sigma}|\phi_{q\sigma}\rangle$,
and it is a matrix called the local occupation matrix. The coefficient
$\kappa_{pq\sigma}$ becomes a matrix as well and is called the curvature
matrix. In the original work of the LOSC (called LOSC1) \cite{2018li215},
$\kappa_{pq\sigma}$ has a similar expression to the GSC case (Eq.~\ref{eq:gsc_kappa})
and it is given as 
\begin{align}
\kappa_{pq\sigma}=\int\frac{\rho_{p\sigma}(\mathbf{r})\rho_{q\sigma}(\mathbf{r'})}{|\mathbf{r}-\mathbf{r'}|}\text{d}\mathbf{r}\text{d}\mathbf{r'}-\frac{2\tau C_{x}}{3}\int[\rho_{p\sigma}(\mathbf{r})]^{\frac{2}{3}}[\rho_{q\sigma}(\mathbf{r})]^{\frac{2}{3}}\text{d}\mathbf{r},\label{eq:kappa1}
\end{align}
in which $\rho_{p\sigma}=|\phi_{p\sigma}|^{2}$ is the density of
the corresponding orbitalet. In the LOSC2 \cite{2020su1535}, the
curvature matrix is modified for better performance in molecules and given as 
\begin{align}
\tilde{\kappa}_{pq\sigma}=\mathrm{erf}(\xi S_{pq\sigma})\sqrt{\kappa_{pp\sigma}\kappa_{qq\sigma}}+\mathrm{erfc}(\xi S_{pq\sigma})\kappa_{pq\sigma},
\end{align}
in which $\xi=8.0$, $S_{pq\sigma}=\int\sqrt{\rho_{p\sigma}(\mathbf{r})\rho_{q\sigma}(\mathbf{r})}\dd\mathbf{r}$,
$\mathrm{erf}(x)$ is the error function and $\mathrm{erfc}(x)$ is
the complementary error function.

\section{Implementation Details}

\begin{figure}[b!]
\centering 
\includegraphics{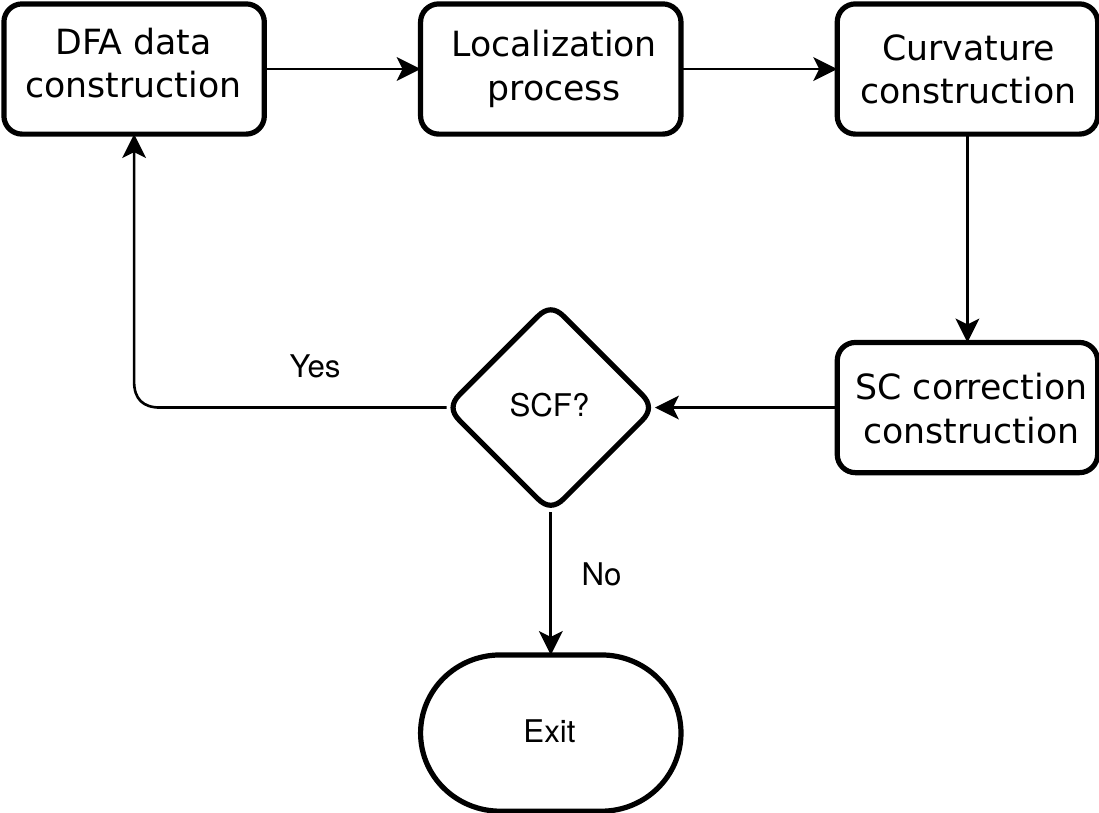}
\caption[The flow chart of scaling correction methods]{The flow chart of scaling correction methods.}
\label{fig:flowchart}
\end{figure}

For the development of \libsc{}, we focus on the implementation
of GSC, LOSC1 and LOSC2 methods at the present stage. Making this
choice is for three reasons: (1) these methods have been shown with
much numerical success; (2) applying calculations from these three
methods only requires a very low computational overhead on top of regular
DFT calculations; (3) these three methods share a similar theoretical
framework and analytical expressions, making their implementations
easy and compact.

To illustrate the design of \libsc{}, we begin with the clarification
for the relations and differences of these three methods in terms
of implementation. First, recall the working flow of the LOSC calculations
as shown in Figure~\ref{fig:flowchart}. A general procedure for
the LOSC calculation in a single self-consistent field (SCF) cycle
includes three steps: (1) conducting the localization; (2) constructing
the curvature matrix; (3) evaluating the corrections to the total
energy and the generalized KS Hamiltonian. The way of conducting steps
(1) and (2), namely, the way of generating orbitalets and constructing
the curvature matrix, differentiates the different versions of the
LOSC. So far, two versions of LOSC, LOSC1 \cite{2018li215} and LOSC2
\cite{2020su1535}, distinguish both in the localization procedure
and the curvature definition. From LOSC1 to LOSC2, both the localization
and the curvature matrices are improved to achieve better performance
in preserving the orbitalet symmetry and degeneracy. Accordingly,
the localization and curvature used in LOSC1 are called \loone{}
and \kappaone{}. Similar names \lotwo{} and \kappatwo{} are
applied to LOSC2. Second, we clarify the connections between the LOSC
and the GSC. To be clear, the GSC \cite{2011zheng026403a} is a special case
in the framework of LOSC (for both LOSC1 and LOSC2), in which the
localization does not take place, yielding orbitalets that are just
canonical orbitals and the curvature matrix is equivalent to the one
defined in Eq.~\ref{eq:gsc_kappa}. Therefore, the general implementation
of LOSC would cover the GSC method with the localization turned off.
To make a clear illustration, we show the relations and comparisons
between GSC, LOSC1 and LOSC2 in Table~\ref{tab:relation}. Only bolded methods 
in the table are supported in \libsc{}.

\begin{table}[htbp]
\centering \caption{Options in the GSC, LOSC1, LOSC2 method and the support from \libsc{}.}
\label{tab:relation} %
\begin{tabular}{|l|c|c|c|}\hline
    \diagbox[width=10em]{Localization \\ Module}{Curvature Matrix\\ Module}&
      \kappazero{} & \kappaone{} & \kappatwo{} \\ \hline
    \lozero{} & \textbf{GSC (v0)}\cite{2011zheng026403a} & \textbf{GSC (v1)} & \textbf{GSC (v2)}\\ \hline
    \loone{}  & N/A                 & L1C1 (LOSC1) \cite{2018li215}     & N/A\\ \hline
    \lotwo{}  & N/A                 & \textbf{L2C1}     & \textbf{L2C2 (LOSC2)}\cite{2020su1535}\\ \hline
    \multicolumn{4}{l}{*The bolded methods are supported in \libsc{}.}
\end{tabular}
\end{table}

Based on the workflow of LOSC and GSC calculations shown in Figure~\ref{fig:flowchart},
we designed \libsc{} as a collection of three modules, namely, the
localization module, the curvature matrix module and the correction
construction module, to provide the essential functionalities for
LOSC and GSC calculations. As the LOSC is a generalized case that
covers the case of the GSC, the implementation of \libsc{} is based
on the expressions from the LOSC. For the localization module, \libsc{}
currently supports only the \lotwo{} \cite{2020su1535}. The Jacob-Sweep
algorithm \cite{1963edmiston464} was implemented to perform the optimization
problem for the localization (see the Supporting Information for details).
For the curvature module, \libsc{} supports three versions, \kappazero{} \cite{2011zheng026403a}, 
\kappaone{} \cite{2018li215} and \kappatwo{} \cite{2020su1535}. 
\kappazero{} is a special case of \kappaone{} and can be called in \libsc{}
by setting the curvature version to 1 and changing the default parameter $\tau$ to 1.00.
Density fitting
\cite{1993vahtras518,1997fruchtl69} is used to evaluate the Coulomb
interaction contribution in the curvature matrix for better efficiency
(see the first term in Eq.~\ref{eq:kappa1} for example). For the
correction construction module, the implementation is straightforward
and based on the analytical expressions (see Eq.~\ref{eq:losc_corr}).

For the details of implementation, we summarize the design choices
in the following:
\begin{itemize}
\item \emph{Languages}: \libsc{} supports three programing languages,
namely \Cpp{}, C and Python. The core of \libsc{} that provides
all the key functionalities of LOSC and GSC methods is implemented
with morden \Cpponeone{} for the consideration of computational
efficiency. The \Cpp{} core of the library is bridged with C and
Python programing languages to provide corresponding interfaces and
avoid code duplications.
\item \emph{Style}: The object-oriented programming (OOP) technique is used
to in the localization module and curvature module to deal with different
versions and avoid code duplications. The functional programming is
used in the correction module and the C code.
\item \emph{Data Structure}: The main data structure in the library is the
matrix object, which is represented by the \texttt{MatrixXd} class
provided by the \texttt{Eigen3} library \cite{eigen3} in low level
\Cpp{} code. The \texttt{Eigen3} library is highly optimized for
the linear algebra manipulations, and heavily used in the SC library
to achieve the best efficiency. The \texttt{MatrixXd} class is mapped
to the \texttt{Numpy} \cite{2011walt30,2020harris362} array object
in Python code and raw C array in C code.
\item \emph{Interface Bridging}: Bridging \Cpp{} library core to the C
code is natural, because these two languages are compatible. Bridging
\Cpp{} library core to the Python code is through the \texttt{pybind11}
library \cite{pybind11}, which provides the supports of binding the
data structure and data types in \Cpp{}, like the class, function
and \texttt{Eigen3} to Python environment. Within the library, most
data are stored in the \Cpp{} library core in memory, and shared
between interfaces to avoid unnecessary data copies. Manipulating
the data from the C or Python interface is efficiently achieved by
directly modifying the corresponding memory blocks through the pointers
or references, which is taken care by the bridging process.
\item \emph{User Interface}: The interfaces for all the functions and classes
in \libsc{} have simple and consistent designs for all the supported
programming languages. The principle is that it mostly takes matrix
objects, rather than complicated and customized class types, as the
main input.
\end{itemize}
With the development of \libsc{}, we can easily implement GSC and
LOSC methods in a quantum chemistry package. Along with \libsc{},
we integrate the library with two popular open-source packages, the
\psipkg{} package \cite{2020smith184108} and the \pyscf{} package
\cite{2018sune1340}. Considering that both packages use the Python
environment for users to conduct calculations, we used the Python
interface of \libsc{} and provided the implementation of LOSC and
GSC methods as Python plugins to both packages, which are the \psilosc{}
plugin for \psipkg{} and the \pyscflosc{} plugin for \pyscf{}.
Note that \libsc{} does not support calculations with symmetry at
the current stage. Therefore, it requires that the symmetry option 
be turned off in both \psipkg{} and \pyscf{} packages to be able
to use \psilosc{} and \pyscflosc{} plugins.

For the \psipkg{} package, the \texttt{Wavefunction} is the main
object that stores all the data from a regular SCF calculation. Therefore,
\psilosc{} communicates with the \psipkg{} package mainly through
the \texttt{Wavefunction} object. Implementing the post-SCF LOSC calculation
within the plugin \psilosc{} is straightforward with assembling
the three steps in a LOSC calculation according to the flowchart in
Figure~\ref{fig:flowchart}. Implementing the SCF-LOSC \cite{2020mei10277}
calculation involves updating the Hamiltonian matrix (or called Fock
matrix in the general SCF cycle within \psipkg{} source code). This
is achieved by overwriting two key functions within \psipkg{} package:
(1) the member function \texttt{Wavefunction.form\_F()} that constructs
the Fock matrix is updated to involve the LOSC effective Hamiltonian
\cite{2020mei10277}; (2) the driver function \texttt{psi4.proc.scf\_wavefunction\_factory()}
that constructs the \texttt{Wavefunction} object for the SCF calculation
is updated to be compatible with the LOSC case. For the density fitting
calculation of curvature matrix in plugin \psilosc{}, the 3-center
integral is constructed block-wise with respect to the fitting basis
index in order to reduce the memory cost.

\begin{singlespace}
\begin{lstlisting}[
                language=python, style=numbers, upquote=true,
                label={lst:psi4_losc},
                caption={Demonstration of using \psilosc{} plugin within
                \psipkg{} package.}
            ]
import psi4      # import the psi4 package
import psi4_losc # import the psi4_losc plugin for LOSC calculations

# Set up a molecule.
# Note the symmetry is turned off in Psi4 with setting C1 symmetry.
mol = psi4.geometry("""
    0 1
    H  0.0  0.0  -0.5
    H  0.0  0.0  0.5
    symmetry c1
    """)

# Conduct the DFA SCF calculation from Psi4.
E_dfa, dfa_wfn = psi4.energy('BLYP', return_wfn=True)

# Configure LOSC calculation settings.
psi4_losc.options.set_param('localizer', 'max_iter', 1000)
# To turn off localization and do GSC, set to be 0.

# Conduct the post-SCF LOSC calculation
E_losc, Orb_losc = psi4_losc.post_scf_losc(
                    psi4_losc.BLYP,  # DFA information
                    dfa_wfn,         # Psi4 wavefunction object
                    window=[-30, 10] # energy window (optional)
                   )

# Conduct the SCF-LOSC calculation
losc_wfn = psi4_losc.scf_losc(
                psi4_losc.BLYP,  # DFA information
                dfa_wfn,         # Psi4 wavefunction object
                window=[-30, 10] # energy window (optional)
           )
\end{lstlisting}
\end{singlespace}

Listing~\ref{lst:psi4_losc} demonstrates the way of using \psilosc{}
plugin within \psipkg{}. As shown in Listing~\ref{lst:psi4_losc},
both the post-SCF and SCF calculations of LOSC is based on a \psipkg{}
\texttt{Wavefunction} object, \texttt{dfa\_wfn}, which can be requested
as the returned value from the \psipkg{} SCF driver function \texttt{psi4.energy()}.
The post-SCF LOSC calculation is conducted by calling the function
\texttt{psi4\_losc.post\_scf\_losc()} provided from the plugin, and
the returned values are the corrected total energy and orbital energies.
The SCF-LOSC calculation is conducted by calling the function \texttt{psi4\_losc.scf\_losc()}
provided by the plugin, and the returned value is a \psipkg{} \texttt{Wavefunction}
object, \texttt{losc\_wfn}, which can be used for further calculations
for property analysis in \psipkg{}.

The implementation of \pyscflosc{} is similar to that of \psilosc{},
with only a few changes to accommodate to features of \pyscf{}.
For the \pyscf{} package, data for a restricted Kohn-Sham (RKS)
or a unrestricted Kohn-Sham (UKS) calculation is stored in a \texttt{pyscf.dft.rks.RKS}
or a \texttt{pyscf.dft.uks.UKS} object respectively. Take the UKS
calculation as an example. All the matrices that are needed for LOSC
calculations can be directly accessed from attributes of the \texttt{pyscf.dft.uks.UKS}
object or be constructed by the corresponding built-in \pyscf{}
functions. For the density fitting calculation of the curvature matrix,
an auxiliary molecule is constructed by the \pyscf{} function \texttt{pyscf.df.addons.make\_auxmol()},
and the 3-center integrals and the 2-center integrals are calculated
directly by using the \pyscf{} built-in functions that interface
with its internal \texttt{Libcint} package\cite{sun2015libcint}.
The implementation of post-SCF LOSC within \pyscflosc{} plugin follows
the flowchart as shown in Figure~\ref{fig:flowchart}, which is similar
to the case of the \psilosc{} plugin. The implementation of SCF-LOSC
provided in the \pyscflosc{} plugin is achieved by overwriting two
member functions, \texttt{get\_fock()} and \texttt{energy\_tot()}
of the \texttt{\pyscf{}.dft.usk.UKS} object, to update the Hamiltonian
matrix and include LOSC contributions within each SCF cycles.

\begin{singlespace}
\begin{lstlisting}[
        language=python, style=numbers, upquote=true,
        label={lst:pyscf_losc},
        caption={Demonstration of using \pyscflosc{} plugin in \pyscf{}
        package.}
    ]
import pyscf        # import the pyscf package
import pyscf_losc   # import pyscf_losc plugin for LOSC calculations

# Initialize a molecule in PySCF
mol = pyscf.gto.Mole()
mol.atom = """
    H  0.0  0.0  -0.5
    H  0.0  0.0  0.5
"""
mol.symmetry = False  # turn off symmetry in PySCF
mol.build()

# Conduct the DFA SCF calculation from PySCF.
mf = pyscf.scf.UKS(mol)
mf.xc = "BLYP"
mf.kernel()

# Configure LOSC calculation settings.
pyscf_losc.options.set_param('localizer', 'max_iter', 1000)
# To turn off localization and do GSC, set to be 0.

# Conduct the post-SCF LOSC calculation
Elosc, Orblosc = pyscf_losc.post_scf_losc(
                       pyscf_losc.BLYP,  # DFA information
                       mf,               # PySCF DFT object
                       window=[-30, 10]  # energy window (optional)
                 )

# Conduct the SCF-LOSC calculation
loscmf = pyscf_losc.scf_losc(
            pyscf_losc.BLYP,  # DFA information
            mf,               # PySCF DFT object
            window=[-30, 10]  # energy window (optional)
         )
\end{lstlisting}
\end{singlespace}

Listing~\ref{lst:pyscf_losc} demonstrates the way of using the \pyscflosc{}
plugin within \pyscf{}. The usage of the \pyscflosc{} plugin is
the same as the \psilosc{} plugin. The only difference is changing
the \texttt{Wavefunction} object used in \psilosc{} to be the \texttt{pyscf.dft.uks.UKS}
object used in \pyscflosc{}.

The source code and documentations for the \libsc{} library and
\psilosc{} and \pyscflosc{} plugins are hosted on Github \cite{libscgithub}.

\section{Applications and Results}

\subsection{Computational Details}

The performance of \libsc{} was first tested by reproducing a series
of calculations that have been done with \qmfd{}\cite{qm4d}. For
more information about molecular structures and reference values,
refer to previous works on the LOSC.\cite{2020mei} \cite{2019mei}\cite{2020su1535}
For G2-1 atomization energy (AE) test set, Hydrocarbon AE test set,
NonHydrocarbon AE test set, SubHydrocarbon AE test set, Radical AE
test set, IP test set, EA test set, HTBH38 reaction barrier (RB) test
set, and NHTBH38 RB test set, both B3LYP and BLYP calculations were
done using 6-311++G(3df, 3pd) basis set with \psipkg{}. Quasiparticle
energies (QE) of a series of systems were calculated using B3LYP functional
and cc-pVTZ basis set with \psipkg{}. To compare the results from
\texttt{psi4\_losc} and from \texttt{pyscf\_losc}, both B3LYP/cc-pVTZ
and BLYP/cc-pVTZ calculations of IP values for polyacetylene (PA)
molecules with 1 to 10 units were done with both \psipkg{} and \pyscf{}.
Photoemission spectra of maleic anhydride were calculated using B3LYP/cc-pVTZ
with \qmfd{}, \psipkg{}, and \pyscf{}. To test the stablity
of the localization procedure, post-LOSC2 calculations of the 4-unit
polyacene molecule were performed starting from random initial $U$ 
matrices with \qmfd{} based
on B3LYP/cc-pVTZ calculations. The aug-cc-pVTZ-RIFIT basis set is used as the density
fitting basis for the construction of curvature matrices for all calculations.
Detailed results are documented in the Supporting Information.

\subsection{Numerical Results}

\begin{table}
\centering \caption{IP values (in eV) of polyacetylene with different unit numbers calculated
from B3LYP and compared with the first IP from RASPT2. The localization
procedure was carried out with the energy window of {[}$-$30, 10{]}.}
\label{tab:pa-b3lyp} %
\begin{tabular}{ccccccccccc}
\toprule 
\multirow{1}{*}{units} & \multirow{1}{*}{Ref} & \multicolumn{3}{c}{B3LYP} & \multicolumn{3}{c}{post-LOSC2} & \multicolumn{3}{c}{SCF-LOSC2}\tabularnewline
\cmidrule(l{0.5em}r{0.5em}){3-5}\cmidrule(l{0.5em}r{0.5em}){6-8}\cmidrule(l{0.5em}r{0.5em}){9-11} &  & \psipkg{} & \qmfd{} & \pyscf{} & \psipkg{} & \qmfd{} & \pyscf{} & \psipkg{} & \qmfd{} & \pyscf{}\tabularnewline
\midrule 
1 & 10.48 & 7.62 & 7.62 & 7.62 & 10.59 & 10.59 & 10.59 & 10.59 & 10.59 & 10.59\tabularnewline
2 & 9.18 & 6.58 & 6.58 & 6.58 & 9.37 & 9.37 & 9.37 & 9.37 & 9.37 & 9.37\tabularnewline
3 & 8.18 & 6.04 & 6.03 & 6.04 & 8.20 & 8.18 & 8.20 & 8.19 & 8.17 & 8.19\tabularnewline
4 & 7.69 & 5.70 & 5.70 & 5.70 & 7.95 & 7.94 & 7.96 & 7.92 & 7.91 & 7.92\tabularnewline
5 & 7.33 & 5.47 & 5.47 & 5.47 & 7.68 & 7.77 & 7.68 & 7.63 & 7.76 & 7.63\tabularnewline
6 & 7.04 & 5.31 & 5.31 & 5.31 & 7.58 & 7.58 & 7.58 & 7.57 & 7.56 & 7.56\tabularnewline
7 & 6.85 & 5.18 & 5.18 & 5.18 & 7.44 & 7.37 & 7.44 & 7.44 & 7.38 & 7.42\tabularnewline
8 & 6.66 & 5.08 & 5.08 & 5.08 & 7.21 & 7.13 & 7.21 & 7.22 & 7.31 & 7.19\tabularnewline
9 & 6.56 & 5.00 & 5.00 & 5.00 & 7.08 & 7.07 & 7.08 & 7.11 & 7.08 & 7.07\tabularnewline
10 & 6.41 & 4.93 & 4.93 & 4.93 & 7.00 & 7.03 & 7.00 & 7.04 & 7.04 & 6.99\tabularnewline
\midrule 
 & MAE & 1.95 & 1.95 & 1.95 & 0.37 & 0.36 & 0.37 & 0.37 & 0.38 & 0.36\tabularnewline
\bottomrule
\end{tabular}
\end{table}

\begin{table}
\centering \caption{MAEs of AE, RB, IP, EA, and QE test sets. Results of AE and RB test
sets are in kcal/mol. Results of IP, EA and QE test sets are in eV.}
\label{tab:maesummary} %
\begin{tabular}{lcccccc}
\toprule 
\multirow{1}{*}{Test set} & \multicolumn{3}{c}{\psipkg{}} & \multicolumn{3}{c}{\qmfd{} \cite{qm4d}}\tabularnewline
\cmidrule(l{0.5em}r{0.5em}){2-4}\cmidrule(l{0.5em}r{0.5em}){5-7} & B3LYP & post-LOSC2 & SCF-LOSC2 & B3LYP & post-LOSC2 & SCF-LOSC2\tabularnewline
\midrule 
\textbf{AE} &  &  &  &  &  & \tabularnewline
G2-1 & 3.80 & 3.80 & 3.80 & 2.45 & 2.45 & 2.45\tabularnewline
NonHydrocarbon & 6.99 & 6.99 & 6.99 & 7.64 & 7.65 & 7.65\tabularnewline
Hydrocarbon & 3.51 & 3.52 & 3.51 & 3.47 & 3.48 & 3.48\tabularnewline
SubHydrocarbon & 2.24 & 2.24 & 2.24 & 2.52 & 2.53 & 2.52\tabularnewline
Radical & 2.31 & 2.30 & 2.30 & 2.26 & 2.26 & 2.26\tabularnewline
\midrule 
\textbf{RB} &  &  &  &  &  & \tabularnewline
HTBH38 & 4.35 & 4.35 & 4.35 & 4.35 & 4.35 & 4.35\tabularnewline
NHTBH38 & 7.17 & 7.17 & 7.17 & 6.38 & 6.38 & 6.38\tabularnewline
\midrule 
IP & 4.52 & 0.63 & 0.64 & 3.19 & 0.35 & 0.35\tabularnewline
\midrule 
EA & 3.47 & 0.48 & 0.47 & 2.57 & 0.44 & 0.44\tabularnewline
\bottomrule
\end{tabular}
\end{table}

\begin{figure}[htbp]
    \centering
    \subfigure[GSC]
    {\includegraphics[width=0.48\linewidth]{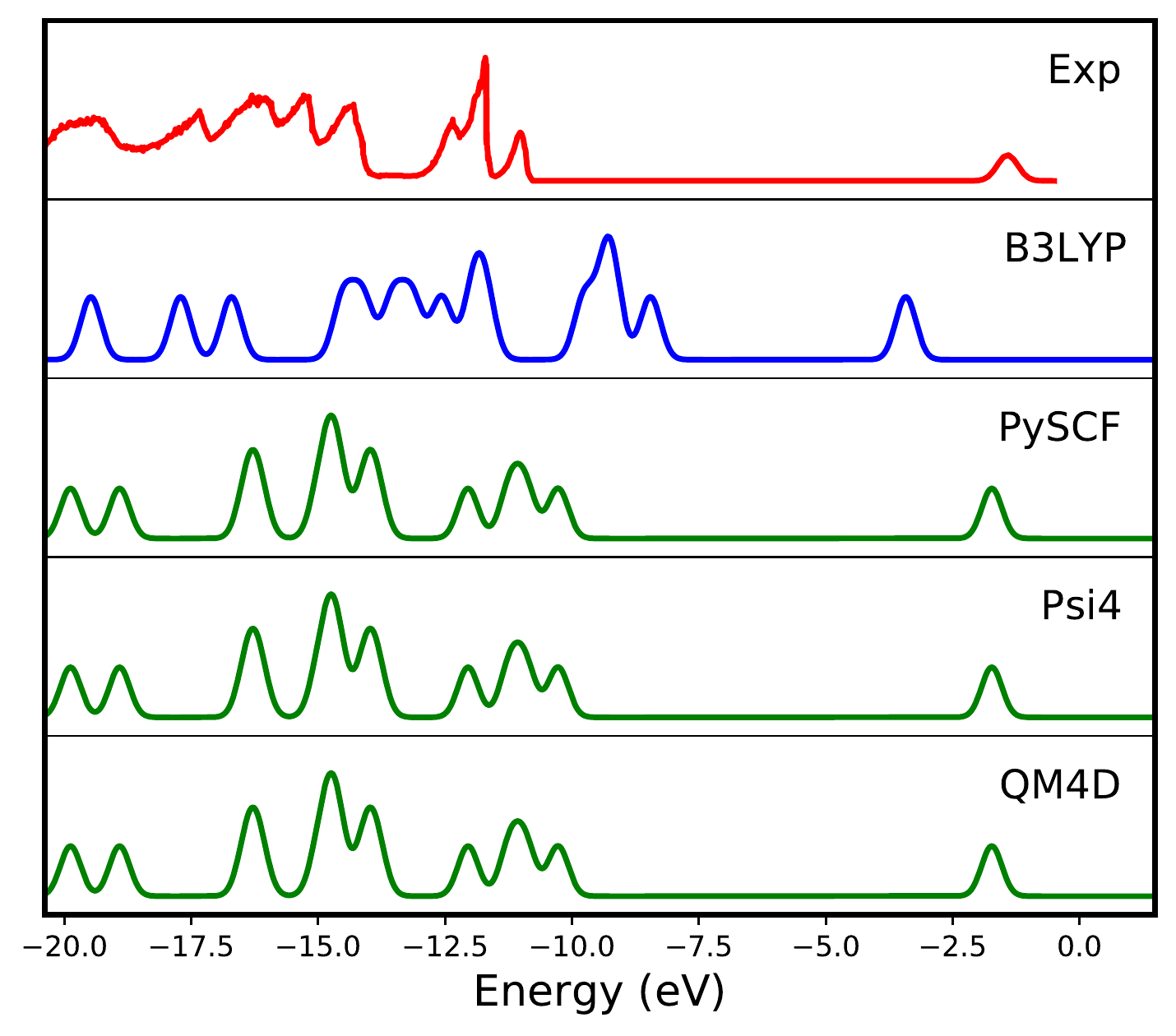}}
    \subfigure[post-LOSC2]
    {\includegraphics[width=0.48\linewidth]{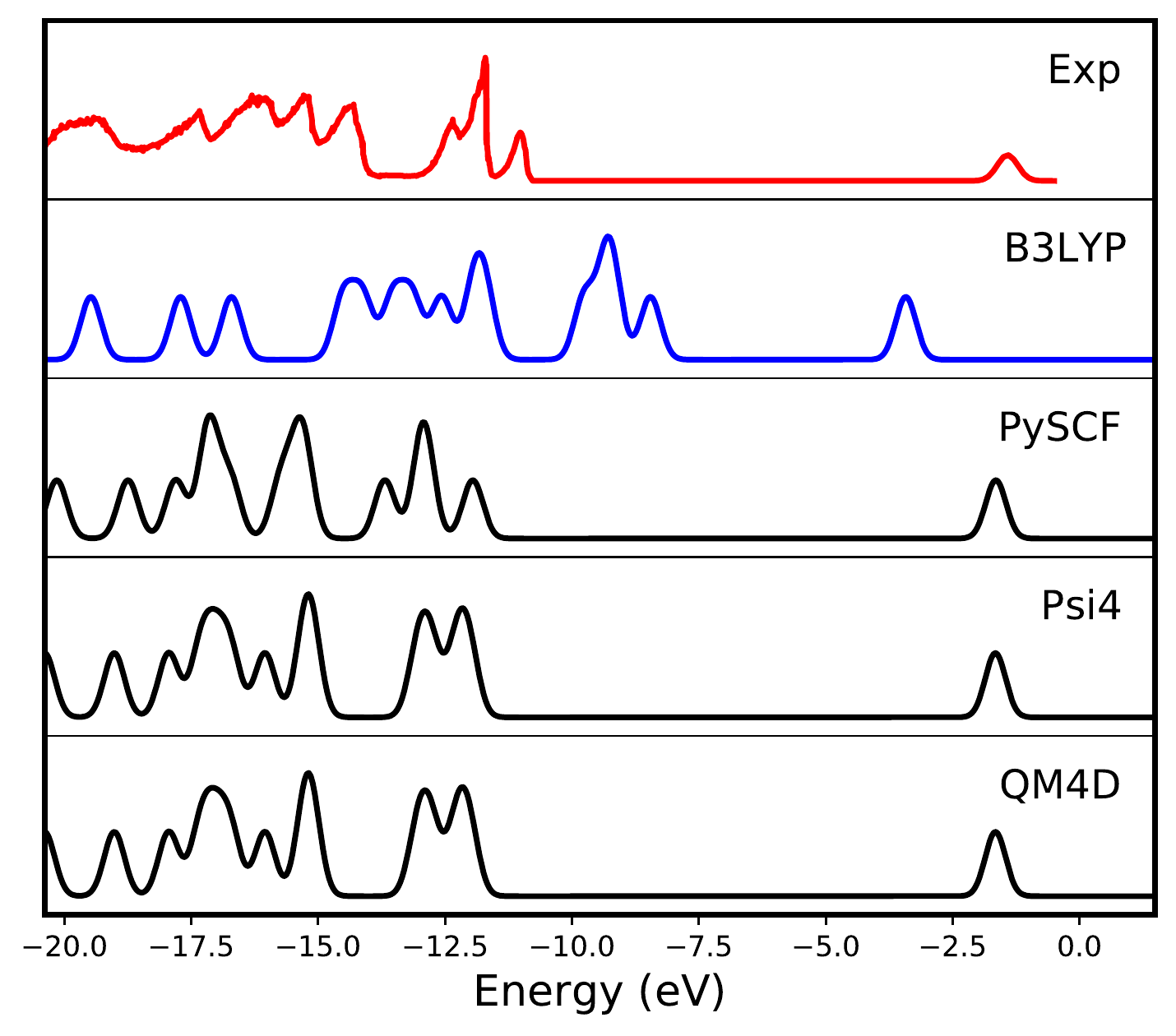}}
    \caption{Photoemission spectrum
    of maleic anhydride calculated from B3LYP, GSC, and post-LOSC2, 
    compared with the experimental result\cite{knight2016accurate}.
    }
    \label{fig:spec}
\end{figure}

To test the implementation of \texttt{psi4\_losc} and \texttt{pyscf\_losc},
IP values of polyacetylene (PA) molecules with 1 to 10 units were
calculated with both \psipkg{} and \pyscf{} and compared with
the results from \qmfd{} as shown in Table~\ref{tab:pa-b3lyp}.
These three software packages give mostly the same results for parent
DFA calculations. The mean absolute errors (MAEs) differ by 0.01 eV
for post-LOSC2 results and 0.02 eV for SCF-LOSC2 results. This shows
that the integration of \libsc{} with both \psipkg{} and \pyscf{}
is correct.

MAEs for the same test sets calculated with \psipkg{} and \qmfd{}
are listed in Table~\ref{tab:maesummary}. The difference between
the numbers is due to the fact that the MAE values are not calculated
with exactly the same systems for each set. We could only get the
results for a subset of the systems because of the SCF convergence
issue of DFT. Detailed results are documented in the Supporting Information.

The photoemission spectrum of maleic anhydride is calculated from
B3LYP, GSC, and post-LOSC2 with all the three software packages. The
results are plotted and compared with both experimental\cite{knight2016accurate}
and B3LYP spectra in Figure~\ref{fig:spec}. For the B3LYP spectrum,
only \qmfd{} result is shown because the three software packages
produce the same result. These spectra show that both GSC and post-LOSC2
greatly correct the behavior of the parent DFA. On the experimental
curve, the peak with the highest energy represents the LUMO energy,
which is about $-$1.3 eV. The second peak from the right represents
the HOMO energy, which should be about $-$11 eV. The HOMO-LUMO gap
is about 10 eV. However, B3LYP gives a HOMO of about $-$8.5 eV and
a LUMO of about $-$3.5 eV, and the HOMO-LUMO gap is about 5 eV, which
is about 5 eV smaller than the experimental one. Both GSC and post-LOSC2
give nearly the same LUMO energy as the experimental one. The GSC
HOMO is about 1 eV higher than the experimental HOMO energy, which
gives a HOMO-LUMO gap about 1 eV smaller than the experimental gap.
The post-LOSC2 HOMO energy is about 1 eV lower than the experimental
value, and the HOMO-LUMO gap is about 1 eV wider than the experimental
gap. Compared with the B3LYP spectrum, GSC and post-LOSC2 spectra
can describe the energy structure of this system better.

\subsection{Multi-minimum Problem with the Localization Procedure}

The spectra shown in Figure~\ref{fig:spec} also indicate that the
minimization of the cost function can have different local solutions.
For GSC calculations, the localization process in Figure~\ref{fig:flowchart}
is skipped, and the SC correction only comes from the curvature. As
shown in Figure~\ref{fig:spec}(a), without the localization process,
the GSC spectra have no noticeable difference. On the other hand,
for the post-LOSC2 calculations, the SC correction comes from both
localization contribution and curvature contribution. With the contribution
from the localization process included, as shown in Figure~\ref{fig:spec}(b),
difference between post-LOSC2 spectra shows up. There is one peak
at around 13 eV in the \pyscf{} spectrum which does not exist in
other two spectra, and the peak at around 15.7 eV which appears in
the \psipkg{} curve and the \qmfd{} curve is missing from the
\pyscf{} curve. These differences, especially the differences between
the \pyscf{} spectrum and the \psipkg{} spectrum, which were calculated
using the same \Cpp{} core code, indicate that the localization
process happened differently for this system.

The problem is not with the implementation of \libsc{}. Repeating
the same calculation with \qmfd{} but starting the localization
process from different initial guesses can also result in different
solutions. Table~\ref{tab:localsln} shows the results of 9 tests,
for which the localization procedure started with 9 random $U$ 
matrices. Seven of the nine tests gave similar HOMO
energies which are around $-$7.19 eV, while the other two HOMO energies
are about $-$6.67 eV. The HOMO energy difference between these two
groups is about 0.5 eV, which is close to the MAE of post-LOSC2 for
the IP test set as shown in Table~\ref{tab:maesummary}. In principle,
the global minimum of the localization cost function, Eq. \ref{eq:localization}
, should be the unique answer. However, as shown in the fifth column
of the table, the cost function can have different values. For a given
system, $\langle r^{2}\rangle$ is a constant, and can be subtracted
from the the cost function. Column 6 lists the relative cost function
values of each test with respect to that of the first test. Tests
7 and 8 have the closest values of the cost function, but the HOMO
energies differ by about 0.5 eV. On the other hand, the multi-minimum
issue does not have significant influence on the total energy. When the 
HOMO energy is about $-$6.67 eV, the total energy is a 
little lower, but the maximum difference is less than 0.0001 Hartree. 
Although tests 8 and 9 give lower total energies, it is not evident that 
they offer better solutions. The cost function but not the total energy 
is the target function of the localization procedure. Unfortunately, the 
values of the cost function differ by about 0.5 a.u.$^2$ for tests 8 and 
9, which means that they give very different local solutions.
\begin{table}
\centering \caption{HOMO energies (in eV), LUMO energies (in eV), HOMO-LUMO gap (in eV),
cost function ($F^{\sigma}-\langle r^{2}\rangle$) values (in a.u.$^2$),
relative cost function values (in a.u.), and total energies (in a.u.$^2$)
from post-LOSC2 calculations with default $\gamma$ (0.707) for the
4-unit polyacene molecule starting from random initial guesses. The
calculations were done with \qmfd{}. The localization procedure
was done with the energy window of {[}$-$30, 10{]}.}
\label{tab:localsln} %
\begin{tabular}{cllllll}
\toprule 
test & $\varepsilon_\text{HOMO}$ & $\varepsilon_\text{LUMO}$ & gap & $F^{\alpha}-\langle r^{2}\rangle$ & $\Delta F^{\alpha}$ & $E_\text{total}$\tabularnewline
\midrule 
1 & $-$7.18545464 & $-$0.76867342 & 6.41678122 & 1106.627980 & 0 & $-$693.4080891\tabularnewline
2 & $-$7.18574992 & $-$0.77368498 & 6.41206494 & 1105.017501 & $-$1.610479 & $-$693.4080853\tabularnewline
3 & $-$7.18638324 & $-$0.77317354 & 6.41320970 & 1106.165982 & $-$0.461998 & $-$693.4080848\tabularnewline
4 & $-$7.18543475 & $-$0.76846903 & 6.41696572 & 1106.060276 & $-$0.567704 & $-$693.4080890\tabularnewline
5 & $-$7.18543353 & $-$0.76846797 & 6.41696556 & 1104.974744 & $-$1.653236 & $-$693.4080884\tabularnewline
6 & $-$7.18575067 & $-$0.77368597 & 6.41206470 & 1104.560970 & $-$2.067010 & $-$693.4080853\tabularnewline
7 & $-$7.18543678 & $-$0.76847004 & 6.41696674 & 1104.501397 & $-$2.126583 & $-$693.4080884\tabularnewline
8 & $-$6.67494957 & $-$0.77259372 & 5.90235585 & 1104.459009 & $-$2.168971 & $-$693.4081312\tabularnewline
9 & $-$6.67494986 & $-$0.77259475 & 5.90235511 & 1106.001484 & $-$0.626496 & $-$693.4081104\tabularnewline
\bottomrule
\end{tabular}
\end{table}

To further investigate the multi-minimum problem, additional post-LOSC2
calculations were performed, with the energy component of the cost
function ignored by setting the value of $\gamma$ in Eq.~\ref{eq:localization}
to 0. The results are shown in Table~\ref{tab:localwd}. With only
the spacial component included in the cost function, the absolute
$\Delta F^{\alpha}$ values are smaller than those in Table~\ref{tab:localsln}.
This suggests that including the energy component in the cost function
can cause the surface of the cost function to be more rugged. The
total energies agree to 0.001 a.u..
This is another evidence that the localization procedure does not
affect the total energy by much.

\begin{table}
\centering \caption{HOMO energies (in eV), LUMO energies (in eV), HOMO-LUMO gap (in eV),
cost function ($F^{\sigma}-\langle r^{2}\rangle$) values (in a.u.$^2$),
relative cost function values (in a.u.$^2$), and total energies (in a.u.)
from post-LOSC2 calculations with $\gamma=0.00$ for the 4-unit polyacene
molecule starting from random initial guesses. The calculations were
done with \qmfd{}. Energy window: {[}$-$30, 10{]}.}
\label{tab:localwd} %
\begin{tabular}{cllllll}
\toprule 
test & $\varepsilon_\text{HOMO}$ & $\varepsilon_\text{LUMO}$ & gap & $F^{\alpha}-\langle r^{2}\rangle$ & $\Delta F^{\alpha}$ & $E_\text{total}$\tabularnewline
\midrule 
1 & $-$8.14696460 & 0.72329534 & 8.87025994 & 1570.28672 & 0 & $-$693.3591227\tabularnewline
2 & $-$8.14507486 & 0.71698648 & 8.86206134 & 1570.75105 & 0.46433 & $-$693.3583495\tabularnewline
3 & $-$8.14299246 & 0.71400910 & 8.85700156 & 1570.53514 & 0.24843 & $-$693.3591227\tabularnewline
4 & $-$8.16482969 & 0.73987472 & 8.90470441 & 1570.33667 & 0.04995 & $-$693.3595938\tabularnewline
5 & $-$8.16672873 & 0.74241664 & 8.90914537 & 1570.54338 & 0.25666 & $-$693.3590521\tabularnewline
6 & $-$8.17631887 & 0.75790963 & 8.93422850 & 1569.94037 & $-$0.34635 & $-$693.3573091\tabularnewline
7 & $-$8.14824556 & 0.72607677 & 8.87432233 & 1570.34364 & 0.05693 & $-$693.3575385\tabularnewline
8 & $-$8.14695585 & 0.72331595 & 8.87027180 & 1570.28789 & 0.00117 & $-$693.3576098\tabularnewline
9 & $-$8.14590040 & 0.71864483 & 8.86454523 & 1570.72905 & 0.44234 & $-$693.3585760\tabularnewline
\bottomrule
\end{tabular}
\end{table}

Data in Tables~\ref{tab:localsln} and \ref{tab:localwd} show that
the energy component can cause the cost function surface to be more
rugged. 
To study the effect of including virtual orbitals, a new set of tests
were done with only all the occupied orbitals included in the localization
procedure. The results are shown in Table~\ref{tab:localocc}. With
$\gamma$ set to zero and no virtual orbital included, the localization
procedure becomes the original Foster-Boys localization\cite{boys1960construction}.
The value of the cost function is very stable. There is no noticeable
difference between the cost function values of these tests, and the
HOMO energies agree to the fifth digits after the decimal point. Thus,
only one solution was found for the Foster-Boys localization. Compared
with Table.~\ref{tab:localwd}, this suggests that including virtual
orbitals in the localization procedure causes the smooth surface of
the original Foster-Boys target function to become more rugged.

We conclude that the localization procedure is very sensitive to the
starting point. Both including the virtual orbitals in the localization
process and adding the energy component in the cost function contribute
to the roughness of the cost function landscape. While multiple patterns
of localization were observed, the total energy is not significantly
affected, and the HOMO energy difference is about the same magnitude
of the MAE of post-LOSC2 for the IP test set. Differences between
post-LOSC2 spectra shown in Figure~\ref{fig:spec} can be attributed
to the slight differences between the converged DFT results from the
three software packages. Currently, it remains a challenge to overcome
this multi-minimum problem.

Fortunately, the multi-minimum problem may not be a serious issue
in practical calculations. Using a random $U$
matrix as the starting point of the localization is just for a testing
purpose. In practical calculations, the initial $U$ matrix
is usually an identity matrix, meaning the use of CO as the initial guess. 
Ten tests, for which the localization
procedure started from an identity $U$ matrix, were performed
on each of three different computers. Detailed data from these tests
are documented in the Supporting Information. The behavior of the localization procedure
is very stable. Results calculated by the same computers including
frontier orbital energies, cost function values, and total energies
are completely identical. The difference between data calculated by
different computers is negligible. The difference between HOMO energies
is about $2\times10^{-5}$ eV. Although currently there is no effective
way to make sure that the global minimum of the cost function is obtained,
the same local minimum can be obtained by keep using the identity
matrix as the initial $U$ matrix. This can also be verified
by the fact that the \psipkg{} results agree with the \qmfd{}
results.

\begin{table}
\centering \caption{HOMO energies (in eV), cost function ($F^{\alpha}-\langle r^{2}\rangle$)
values (in a.u.$^2$), relative cost function values
(in a.u.$^2$), and total energies (in a.u.) from post-LOSC2 calculations
with $\gamma=0.00$ for the 4-unit polyacene molecule starting from
random initial guesses of the $U$ matrix. The calculations were done
with \qmfd{}. Only occupied orbitals were included in the localization
procedure.}
\label{tab:localocc} %
\begin{tabular}{cllll}
\toprule 
test & $\varepsilon_\text{HOMO}$ & $F^{\alpha}-\langle r^{2}\rangle$ & $\Delta F^{\alpha}$ & $E_\text{total}$\tabularnewline
\midrule 
1 & $-$9.08660137 & 289.22246 & 0 & $-$693.4083449\tabularnewline
2 & $-$9.08660080 & 289.22246 & 0 & $-$693.4083449\tabularnewline
3 & $-$9.08660401 & 289.22246 & 0 & $-$693.4083449\tabularnewline
4 & $-$9.08660259 & 289.22246 & 0 & $-$693.4083449\tabularnewline
5 & $-$9.08660203 & 289.22246 & 0 & $-$693.4083449\tabularnewline
6 & $-$9.08660303 & 289.22246 & 0 & $-$693.4083449\tabularnewline
7 & $-$9.08660383 & 289.22246 & 0 & $-$693.4083449\tabularnewline
8 & $-$9.08660194 & 289.22246 & 0 & $-$693.4083449\tabularnewline
9 & $-$9.08660198 & 289.22246 & 0 & $-$693.4083449\tabularnewline
\bottomrule
\end{tabular}
\end{table}

\section{Conclusion}

In summary, we developed a reliable, flexible and open-source library
\libsc{} for the scaling correction methods, which supports the
GSC and LOSC methods at the present stage. The consistent and simple
interfaces to multiple programming languages, including C, \Cpp{}
and Python, are carefully designed to be user-friendly. We also applied
\libsc{} in two open-source quantum chemistry packages, \psipkg{}
and \pyscf{}. With the distribution of \libsc{}
and its implementation, the scaling correction methods should be available
for broader applications.
\begin{acknowledgement}
Y. M, J.Y. and Z.C. acknowledge the support from the National Institute
of General Medical Sciences of the National Institutes of Health under
award number R01-GM061870. W.Y. acknowledges the support from the
National Science Foundation (grant no. CHE-1900338). Y. M. was also
supported by the Shaffer-Hunnicutt Fellowship and Z.C. by the Kathleen
Zielik Fellowship from Duke University. N.Q.S. acknowledges the support
from the National Natural Science Foundation of China (grant no. 22073049),
\end{acknowledgement}
\begin{suppinfo}
Supporting Information Available: computational details and numerical
results.
\end{suppinfo}
\bibliography{extra,ref_merged,python}

\begin{figure*}
\includegraphics[scale=0.3]{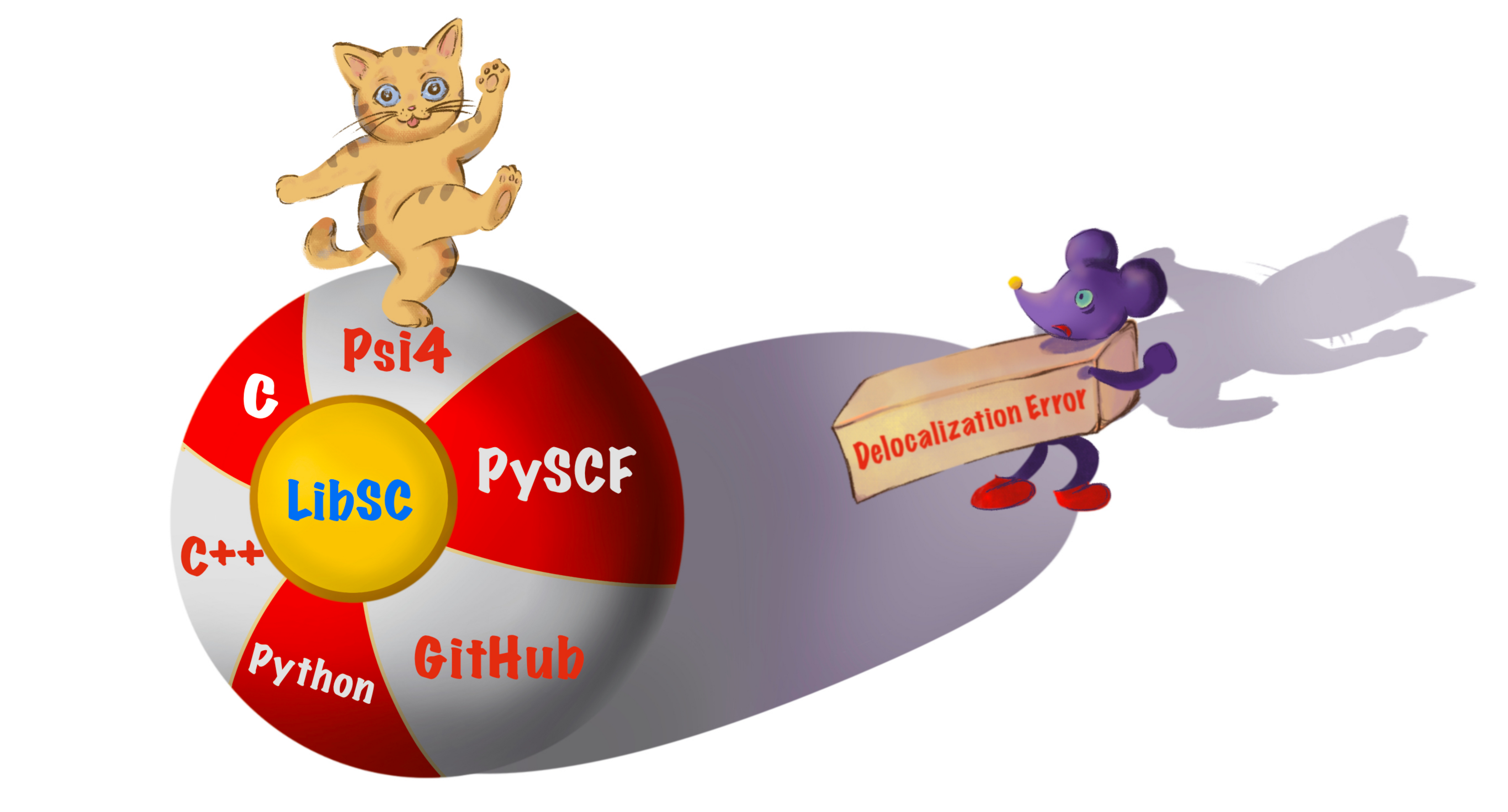}
\caption{For Table of Contents Only}
\end{figure*}

\end{document}